\documentclass[english,prl,twocolumn,showpacs,
superscriptaddress,floatfix,longbibliography]{revtex4-1}

\usepackage[T1]{fontenc}
\usepackage[latin9]{inputenc}
\usepackage{amsmath,amssymb,amstext}
\usepackage{xcolor}
\usepackage{graphicx}
\usepackage{esint}
\usepackage{babel}

\makeatletter
\let\c@author\relax
\makeatother

\usepackage{bm,bbold}
\usepackage{color}
\usepackage{subcaption}
\usepackage{dsfont}
\usepackage[labelfont=bf, justification=justified]{caption}

\definecolor{TODO-color}{HTML}{540682}

\definecolor{COMMENT-color}{HTML}{540685}

\begin{document}
\title{Spectral Properties and Quantum Phase Transitions in Superconducting
  Junctions with a Ferromagnetic Link}
  
\author{M. Rouco}
\email{mikel.rouco@ehu.eus} \affiliation{Centro de F\'isica de Materiales
  (CFM-MPC), Centro Mixto CSIC-UPV/EHU, Manuel de Lardizabal 5, E-20018 San
  Sebastian, Spain}

\author{I. V. Tokatly} 
\email{ilya.tokatly@ehu.es} \affiliation{Nano-Bio Spectroscopy group, Departamento
  F\'isica de Materiales, Universidad del Pa\'is Vasco, Av. Tolosa 72, E-20018
  San Sebasti\'an, Spain} \affiliation{IKERBASQUE, Basque Foundation for
  Science, E48011 Bilbao, Spain} \affiliation{Donostia International Physics Center
  (DIPC),Manuel de Lardizabal 4, E-20018 San Sebastian, Spain}

\author{F. S. Bergeret}
\email{sebastian_bergeret@ehu.eus} \affiliation{Centro de F\'isica de Materiales
  (CFM-MPC), Centro Mixto CSIC-UPV/EHU, Manuel de Lardizabal 5, E-20018 San
  Sebastian, Spain} \affiliation{Donostia International Physics Center
  (DIPC),Manuel de Lardizabal 4, E-20018 San Sebastian, Spain}

\begin{abstract}
  We study theoretically the spectral and transport properties of a
  superconducting wire with a magnetic defect. We start by modelling the system
  as a one dimensional magnetic Josephson junction and derive the equation
  determining the full subgap spectrum in terms of the normal-state transfer
  matrix for arbitrary length and exchange field of the magnetic region.  We
  demonstrate that the quantum phase transition predicted for a short-range
  magnetic impurity, and associated with a change of the total spin of the
  system, also occurs in junctions of finite length. Specifically, we find that
  the total spin changes discontinuously by integer jumps when bounds states
  cross the Fermi level. The spin can be calculated by using a generalization of
  Friedel sum rule for the superconducting state, which we also derive. With
  these tools, we analyze the subgap spectrum of a junction with the length of
  the magnetic region smaller than the superconducting coherence length and
  demonstrate how phase transitions also manifest as change of the sign of the
  supercurrent.
\end{abstract}

\maketitle

\section{Introduction}
\label{sec:intro}
The study of Josephson magnetic junctions and magnetic impurities in
superconductors has attracted a great deal of attention in the past years. The
research is mainly motivated by the search of a topological superconducting
state in magnetic impurity chains and clusters embedded in a superconductor
\cite{yazdani-1997-probing,franke-2011-competition,nadj-2014-observation,heinrich-2018-single,choi-2018-influence}.
In this context it is essential to understand the spectral properties around the
magnetic region. In a quasi one-dimensional setup this is equivalent to study
the spectrum of a superconductor-ferromagnet-superconductor (SFS) junction.

\begin{figure}[t!]
  \centering
  \includegraphics[width=\linewidth]{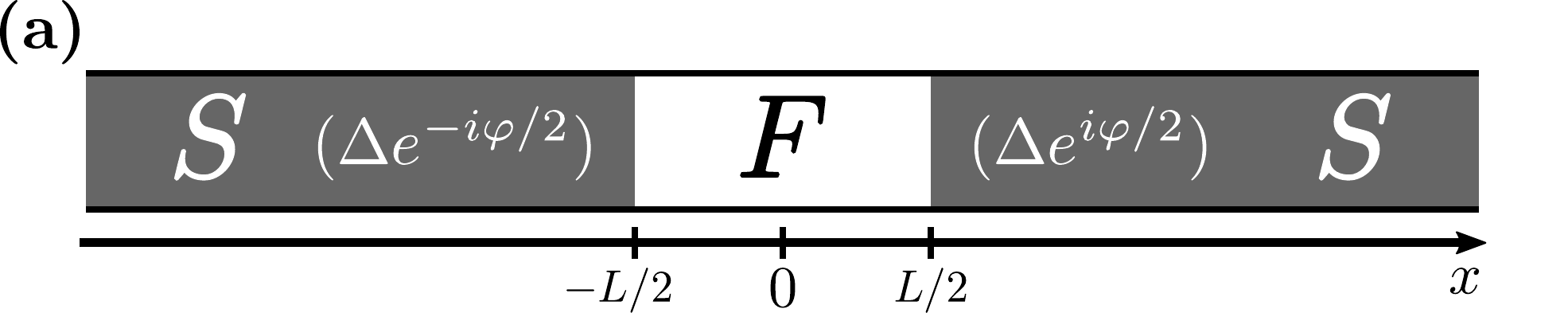}\vspace{0.5em}
  \includegraphics[width=\linewidth]{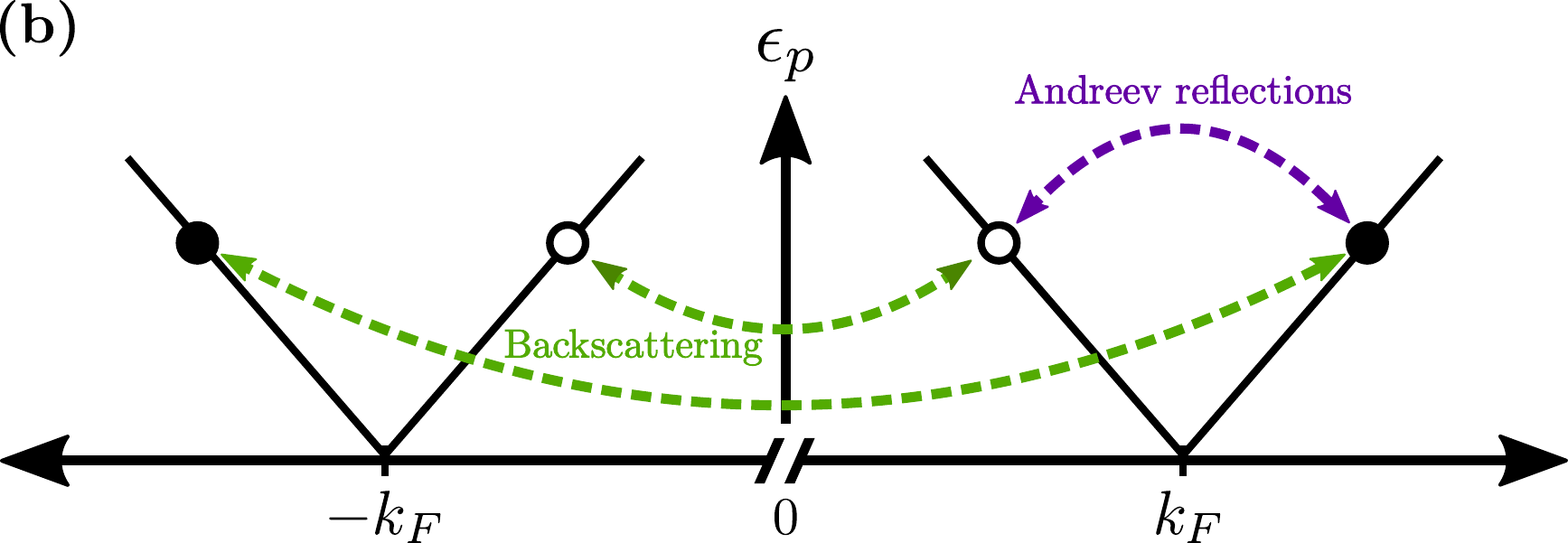}
  \caption{(a) The SFS junction under consideration. (b) Possible processes
    taking place in a Josephson junction: In the absence of normal reflection
    only Andreev reflections within the same propagation valley can occur
    (purple line). If normal backscattering is present valleys at $\pm k_F$ are
    coupled (green lines).}
  \label{fig:sketches}
\end{figure}

Ballistic SFS junctions have been widely explored in the past, mainly in two
limiting cases. One of them is the semiclassical limit, in which the Fermi
energy, $\mu$, is assummed to be much larger than any other energy involved in
the system, including the superconducting gap, $\Delta$, and the Zeeman
splitting, $h$
\cite{golubov_kupriyanov.2004,buzdin.2005_RMP,Bergeret2005,konschelle-2016-semiclassical,
  konschelle-2016-ballistic}. In this limit, one can directly apply the
Bohr-Sommerfeld semiclassical quantization condition
\cite{duncan2002semiclassical} and demonstrate that, in the absence of interface
barriers, the spectrum consists of two double-degenerate Andreev bound states
with opposite energies. This degeneracy of the bound states reflects the
degeneracy of the $\pm k_F$ Fermi momentum valleys, which remain uncoupled in
the absence of normal reflection, as schematically showed in
Fig~\ref{fig:sketches}b.

The second widely studied limiting case is when the spin-splitting field is very
large, $h\gg\mu$, and concentrated in a region much smaller than $k_F^{-1}$
\cite{yu-1965-bound,shiba-1968-classical,rusinov-1968-superconductivity,costa-2018-connection}. This
has been described as $\delta$-like magnetic impurity that strongly couples both
propagation directions to form two nondegenerate bound states within the gap
with opposite energies.  These states, known as the Yu-Shiba-Rusinov (YSR)
states, may cross the Fermi level at a certain strength of the exchange energy.
At this crossing, the system undergoes a quantum phase transition~(QPT)
\cite{sakurai-1970-comments,balatsky-2006-impurity} that has been widely studied
within the $\delta$-like impurity model.  However, the discussion of whether
such a QPT may take place beyond the impurity model is an open question.  To
address it, one needs to understand how these two known limiting cases are
connected.

The goal of this work is twofold. On the one hand, we derive a general equation,
Eq~\eqref{eq:quant-condition}, that determines the subgap spectrum of a
one-dimensional junction in terms of the normal-state transfer matrix for an
arbitrary spin-dependent potential describing the F region, assuming that
$\Delta=0$ within F.  For the particular case of a collinear (unidirectional)
magnetization in the F region, we derive a generalized Friedel sum rule,
Eq~\eqref{eq:magnetization-zero-temp}, adapted for the superconducting
state. This rule states that every time a bound state crosses the Fermi energy,
the total electronic spin changes by the amount of $\hbar/2$. Importantly, this
sum rule is valid not only for one-dimensional systems, but applies universally
to any dimension, size and shape of a localized magnetic region.

On the other hand, in Sec. \ref{sec:model}, we use these findings to provide a
complete analysis of the subgap spectrum of a ballistic one-dimensional SFS
junction for and arbitrary homogenoeus exchange field $h$.  We focus on the
short junction regime, where the ferromagnetic region is shorter than the
superconducting coherence length, $\xi$. In this case, the presence of a
superconducting gap in the ferromagnet due to the proximity effect has no effect
on the subgap spectral properties of the junction, so we set $\Delta=0$ in F.
For this system we obtain the normal-state transfer matrix and, from it, we
determine all spectral properties of the system from the central expression
Eq~\eqref{eq:quant-condition}. We recover the well-established limiting cases,
i.e. delta-like and semiclassical magnetic region, but also the subgap spectrum
for all intermediate situations.  We identify the values of $h$ and $L$ at which
zero-energy crossings of bound states occur.  As in the YSR case, these
crossings are associated with a QPT, which manifests as a change of the total
electronic spin of the system, in accordance with the sum rule derived in
section \ref{sec:spin-counting}. We finally demonstrate that this change of the
total spin at the QPT is associated with the change of sign of the supercurrent
in the SFS junction or, equivalently, to a change of the ground state phase
difference between the superconductors from $0$ to $\pi$.

\section{Model and General Properties}
\label{sec:general}

We consider a one-dimensional geometry consisting of a superconducting wire interrupted
by a ferromagnetic region, as sketched in Fig~\ref{fig:sketches}a. The  Bogoliubov-De Gennes (BdG) Hamiltonian of the system reads
\begin{equation}
  \label{eq:system-hamiltonian}
  \hat H (x) = \left(
    \begin{array}{cc}
      -\frac{\hbar^2 \partial_x^2}{2m} - \mu - V(x) & \Delta(x) \\
      \Delta^*(x) & \frac{\hbar^2 \partial_x^2}{2m} + \mu + \bar{V}(x)
    \end{array}\right)\; .
\end{equation}
Here $\mu$ is the chemical potential, $\Delta(x)$ is the superconducting gap
that is only finite on the S electrodes,
$\Delta(|x|>L/2) = |\Delta| e^{\pm i\varphi/2}$, the length of the F region is
labeled by $L$ and $\pm \varphi/2$ is the superconducting phase, where the plus
(minus) sign stands for the right (left) superconductor. The potential
$V(x) = V_0(x) + \boldsymbol{h}(x)\cdot\boldsymbol{\sigma}$ is only finite, but
arbitrary, within the region $|x|<L/2$ and it consists of an scalar component
$V_0$ and a spin-dependent one $\boldsymbol{h}(x)\cdot\boldsymbol{\sigma}$. The
"bar" denotes time-reverse conjugation such that
$\bar V=\hat\sigma_yV^*\hat\sigma_y$.

We focus on the subgap spectra, $\epsilon < |\Delta|$, which determines the main
transport features at zero voltage and low temperatures. For such energies, the
decaying wavefunctions into the left (L) and right (R) superconducting leads for
each spin component read
\begin{equation}
  \label{eq:wavefunction-L}
  \begin{array}{ll}
    \Psi^\sigma_L (x<\tfrac{-L}{2}) = & e^{x/\xi}
                                        \Bigg[  \Bigg(
                                        \begin{array}{c}
                                          A_L^\sigma \\ A_L^\sigma e^{i\alpha} e^{-i\varphi/2}
                                        \end{array}\Bigg) e^{ik_F x}  \
    \\[1em]
                                      & +  \left(
                                        \begin{array}{c}
                                          B_L^\sigma \\ B_L^\sigma e^{-i\alpha} e^{-i\varphi/2}
                                        \end{array}\right) e^{-ik_F x} \Bigg],
  \end{array}
\end{equation}

\begin{equation}
  \label{eq:wavefunction-R}
  \begin{array}{ll}
    \Psi^\sigma_R (x>\tfrac{L}{2}) = & e^{-x/\xi}
                                       \Bigg[  \Bigg(
                                       \begin{array}{c}
                                         A_R^\sigma \\ A_R^\sigma e^{-i\alpha} e^{i\varphi/2}
                                       \end{array}\Bigg) e^{ik_F x}\
    \\[1em]
                                      & +  \left(
                                        \begin{array}{c}
                                          B_R^\sigma \\ B_R^\sigma e^{i\alpha} e^{i\varphi/2}
                                        \end{array}\right) e^{-ik_F x} \Bigg],
  \end{array}
\end{equation}
where the upper (lower) element of the Nambu spinors stand for electrons
(holes), the index $\sigma=\pm$ labels components of the spin spinor,
$\xi = \hbar v_F/\sqrt{\Delta^2-\epsilon^2}$ is the decaying length of the
wavefunction into the superconductor and $k_F$ and $v_F$ stand for the Fermi
wavenumber and the Fermi velocity respectively. The quantity $\alpha$ is the
phase associated with each Andreev reflection at the S/F interface and it is
given by  $\cos \alpha = \frac{\epsilon}{\Delta}$.

The coefficients $A_{L(R)}^\sigma$ and $B_{L(R)}^\sigma$ in Eqs
\eqref{eq:wavefunction-L} and \eqref{eq:wavefunction-R} are the constants of
integration at the left (right) superconductor for the quasiparticles consisting
of right moving (those multiplied by $e^{ik_Fx}$) and left moving (those
multiplied by $e^{-ik_Fx}$) electrons respectively.  At this stage it is
convenient to define the four vectors
$\textbf{C}_{L(R)} \equiv (A_{L(R)}^+, B_{L(R)}^+, A_{L(R)}^-, B_{L(R)}^-)^T$
for the left (right) superconductor.

The wave functions on  opposite sides of the F region are connected via the
normal state electronic T-matrix, $\check T$
\begin{equation}
  \label{eq:L-to-R-connection-electrons}
  \textbf{C}_R = \check T  \textbf{C}_L
\end{equation}
for the electrons and
\begin{equation}
  \label{eq:L-to-R-connection-holes}
  \textbf{C}_R = e^{-i\varphi} e^{i\hat\alpha}
  \check{\overline T} e^{i\check\alpha} \textbf{C}_L
\end{equation}
for the holes. In Eqs \eqref{eq:L-to-R-connection-electrons} and
\eqref{eq:L-to-R-connection-holes}, $e^{i\hat\alpha}$ is a diagonal matrix with
elements $[e^{i\alpha}, e^{i\alpha},e^{-i\alpha}, e^{-i\alpha}]$.  Notice that
time conjugation also implies to change the sign of the quasiparticle energy
($\epsilon \rightarrow -\epsilon$), so that
$\check{\overline T}(\epsilon)=\hat\sigma_y\check{T}^*(-\epsilon)\hat\sigma_y$.

After substitution of $\mathbf{C}_R$ from Eq
\eqref{eq:L-to-R-connection-electrons} into Eq
\eqref{eq:L-to-R-connection-holes} and multiplication by $\check T^{-1} $ from
the left one obtains a homogeneous equation for $\mathbf{C}_L$ that leads  to the
condition determining the bound states:
\begin{equation}
  \label{eq:quant-condition}
  \det\bigg(
    e^{i\varphi} - 
    \check T^{-1} e^{i\hat\alpha} \check{\overline T} e^{i\hat\alpha} 
  \bigg)=0\; .
\end{equation}
This expression is a generalization of Beenakker's equation for the Andreev
spectrum of a SNS junctions derived from the scattering
matrix~\cite{beenakker-1991-universal}.  The second term inside the determinant
describes an ``Andreev loop''. Namely, from right to left, first an electron
from F is Andreev reflected as a hole at one F/S interface. The hole propagates
to the opposite interface and it is converted again into an electron via the
Andreev reflection. The electron is finally transferred back to the origin.
After this cycle, the wavefunction accumulates a phase equal to $\varphi$.

\subsection{Sum rule for spin in gapped systems}
\label{sec:spin-counting}

Before using Eq~\eqref{eq:quant-condition} to calculate the subgap spectrum in a
one-dimensional geometry, we can anticipate changes of the total spin of the
system associated with bound states crossing the Fermi level. Note that the
derivation presented here is valid for any dimension, so that the result that we
obtain is not restricted to the one dimensional problem described by Eq
\eqref{eq:system-hamiltonian}.

We start by considering the retarded Green's function (GF) for the BdG
equations,
\begin{equation}
  \label{eq:GF-complete}
  \hat G^R(\epsilon) = (\epsilon - \hat H_0 - \hat V + i0^+)^{-1},
\end{equation}
where $\hat H_0$ is the unperturbed BdG Hamiltonian of the system and $\hat V$
is a general perturbation potential operator, as introduced in Eq~\eqref{eq:quant-condition} . The component $i=\{x,y,z\}$ of the
total electronic spin is given by
\begin{equation}
  \label{eq:magnetization-general}
  S_i = -\frac{\hbar}{4\pi} \int_{-\infty}^{\infty} d\epsilon f_F(\epsilon) \text{Im} \bigg[
  \text{Tr} \Big\{{\hat\sigma}_i\hat G^R(\epsilon) \Big\} \bigg],
\end{equation}
where $f_F(\epsilon)=(e^{\epsilon/k_BT}+1)^{-1}$ is the Fermi distribution
function, ${\hat\sigma}_i$ is the $i$-th Pauli matrix, and the trace runs over
the whole coordinate$\times$Nambu$\times$spin space.

The full GF in Eq~\eqref{eq:magnetization-general} can be also written in
terms of the unperturbed GF, $\hat G_0$, and the  potential $\hat V$ via
Dyson's  equation, $\hat G^R = \hat G_0^R + \hat G_0^R \hat V \hat G^R$. Solving
it for $\hat G^R$ and substituting it back into the right hand side, we obtain
the expression determining the exact  $\hat G^R$
\begin{equation}
  \label{eq:GF-complete-Dynes}
  \hat G^R = \hat G_0^R + \hat G_0^R \hat V (I - \hat G_0^R \hat V)^{-1} \hat G_0^R.
\end{equation}

As the total spin of the unperturbed system is zero, only the second term of
$\hat G^R$ in Eq~\eqref{eq:GF-complete-Dynes} contributes to  the trace in
Eq~\eqref{eq:magnetization-general}. 

Let us now assume that $\hat V$ is an energy independent local perturbation, and
its magnetic part is collinear with $z$-axis (i.e. it commutes with
$\hat\sigma_z$). Noticing that
$\big(\hat G_0^R\big)^2 = - \frac{d \hat G_0^R}{d\epsilon}$, one can use the
cyclic property of trace to obtain from Eq~\eqref{eq:magnetization-general} the
$z$-component of the total spin:
\begin{equation}
  \label{eq:magnetization}
  S =\frac{\hbar}{2} \int_{-\infty}^{\infty} \frac{d\epsilon}{2\pi}
  f_F(\epsilon)\frac{d}{d\epsilon}
  \big[\delta_{-}(\epsilon)-\delta_{+}(\epsilon)\big]\; ,
\end{equation}
where
$\delta_{\sigma}(\epsilon)=\text{Im}\ln \big[\det(I - \hat G_0^R(\epsilon)
V_\sigma)\big]$ is a generalized phase shift. Notice that $\hat V$ can have any
spatial distribution and that the determinant inside the logarithm is the
quantization condition coming from the Lippmann-Schwinger equation. In
particular, zeros of this determinant determine the spectrum of the bound
states. Therefore, in a one-dimensional case, it has to be proportional to the
left hand side of Eq~\eqref{eq:quant-condition}. At zero temperature ($T=0$),
Eq~\eqref{eq:magnetization} becomes especially simple,
\begin{equation}
  \label{eq:magnetization-zero-temp}
  2S/\hbar=\frac{1}{2\pi}\big[\delta_{-}(0)-\delta_{+}(0)\big].
\end{equation}

This result is analogous to the well-known Friedel sum rule that relates the
charge/spin induced by a local perturbation to the phase shifts at the Fermi
level.

The important feature of the superconducting state is its gap at the Fermi level
($\epsilon=0$), where the unperturbed Green's function is real (and, therefore,
$\det(I - \sigma \hat G_0^R(0) V)$ is real too). Thus, $\delta_{\sigma}(0)/\pi$
can only take integer values, which will only change discontinuously by $\pm 1$
when a spin polarized bound state crosses the middle of the gap, as the
determinant changes its sign. The electron-hole symmetry requires that the
spin-up/down polarized states cross zero simultaneously while moving in opposite
directions. As a result, at every crossing event the normalized spin $2S/\hbar$
jumps by one \footnote{The stepwise process of the spin polarization that
follows from our phase-shift arguments agree with the picture of
Ref.~\cite{balatsky-2006-impurity} based of the analysis of the spin structure
of the many-body BCS wavefunction in the $\delta$-like impurity case. However,
it should be note that the result of this section is valid for any energy
independent local perturbation potential $\hat V$ acting on a system with a
gap at the Fermi energy and Green's function $\hat G_0$, as long as $\hat V$
commutes both with $\hat G_0$ and $\hat \sigma_z$.}.

In the above derivation we only assume that the perturbation $\hat{V}$ is
localized in space and has a collinear magnetic structure. Therefore our sum
rule relating the total induced spin to the in-gap spectrum applies to any
dimension and any size and shape of a finite magnetic region. For example, it
can be directly used to analyze the behavior of the total spin in a magnetic
chain on top of a superconductor, as the one studied in
Ref~\onlinecite{bjorson-2017-superconducting}.

\begin{figure*}[t!]
  \includegraphics[width=\textwidth]{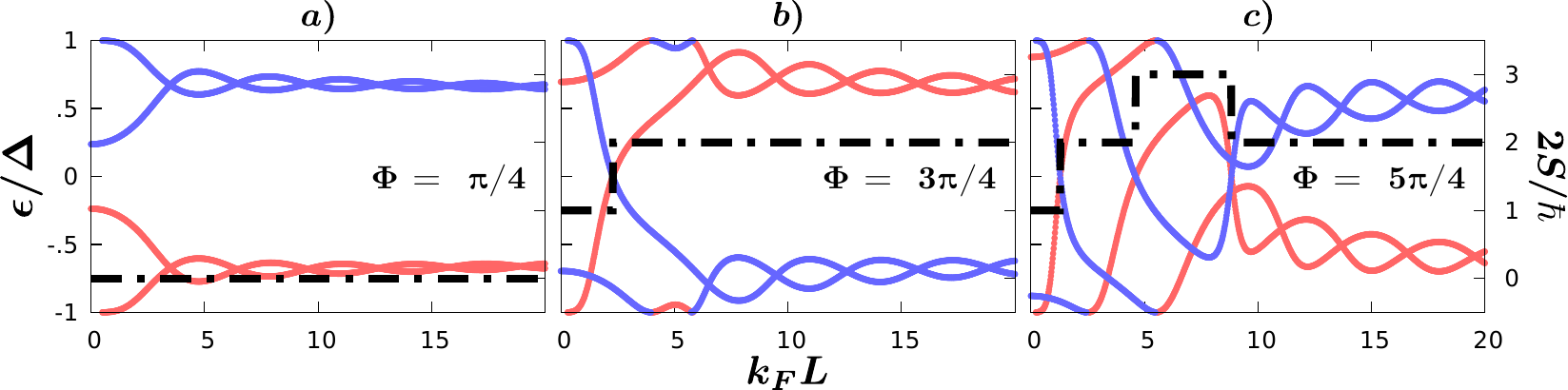}
  \caption{Energy of the bound states (solid lines) and the total spin
    $2S/\hbar$ (dashed line) of a SFS junction as a function of $L$ for three
    different values of $\Phi$, and $\mu / \Delta = 100$. Red and blue colors
    correspond to spin projections of the electronic states.}
  \label{fig:bound-states}
\end{figure*}

\section{ One-dimensional SFS junction}
\label{sec:model}

We now apply the results of previous section to compute the spectral properties
of a one-dimensional SFS junction.  We assume that the scattering F region is
described by the potential $\hat V(x) = h\hat\sigma_z$ for $|x| < L/2$.  In such
a case, the T-matrix  in Eq~\eqref{eq:quant-condition} has a
block-diagonal structure in spin-space,
\begin{equation}
  \label{eq:T-matrix-block-structure}
  \check T = \left(
    \begin{array}{cc}
      \hat T^+ & 0\\
      0 & \hat T^-
    \end{array} \right)
  =
  \left(
    \begin{array}{cccc}
      T^+_{++} & T^+_{+-} & 0 & 0 \\
      T^+_{-+} & T^+_{--} & 0 & 0 \\
      0 & 0 & T^-_{++} & T^-_{+-}\\
      0 & 0 & T^-_{+-} & T^-_{--}
    \end{array}\right)\; .
\end{equation}
This considerably simplifies the problem, since we only need to calculate the
normal state transfer for each spin orientation, $\sigma = \pm$, separately (see
Appendix \ref{sec:app-T-matrix} for details).  The elements of $\check T$ in
Eq~\eqref{eq:T-matrix-block-structure} read
\begin{align}
  \label{eq:T-matrix-elements}
  &T_{++}^{\sigma} = \bigg[
    \cos\Big( q_{\sigma} L \Big) +
    \frac{i}{2}\frac{{q_{\sigma}}^2 + {q_0}^2}{q_{\sigma} q_0}
    \sin \Big(q_{\sigma} L\Big)
    \bigg] e^{-iq_0 L}, \nonumber\\
  &T_{+-}^{\sigma} = \frac{i}{2}\frac{{q_{\sigma}}^2 - {q_0}^2}{q_{\sigma} q_0}
    \sin \Big(q_{\sigma} L\Big)\; ,
\end{align}
where $q_\sigma(\epsilon) = k_F \sqrt{1+\tfrac{\epsilon + \sigma h}{\mu}}$ and
$q_0(\epsilon) = k_F \sqrt{1 + \tfrac{\epsilon}{\mu}}$ are the momentum of the
electron in the ferromagnet and the normal metal respectively and $\mu$ stands
for the Fermi energy. Due to the symmetry of the problem, one can verify that
other components of $\hat T^\sigma$ are related to the ones in
Eq~\eqref{eq:T-matrix-elements} by complex conjugation,
$T_{--}^\sigma = \big(T_{++}^\sigma\big)^*$ and
$T_{+-}^\sigma = \big(T_{-+}^\sigma\big)^*$.  The diagonal terms,
${T}_{aa}^\sigma$, describe a direct transmission (forward scattering) within
one valley, whereas the off-diagonal terms represent backscattering events that
couple the opposite valleys at $\pm k_F$ (see Fig~\ref{fig:sketches}b).

The solution of Eq~\eqref{eq:quant-condition}, after substitution of
Eq~\eqref{eq:T-matrix-elements} in it, determines the full subgap spectrum of
the SFS junction. For analytic results, we focus on the semiclassical limit
where $\mu$ is the largest energy, so that $\epsilon, \Delta,h \ll \mu$.  In
this case the quassiparticle momenta in the F and S regions are approximated by
$q_\sigma(\epsilon) \approx k_F + \tfrac{\epsilon + \sigma h}{\hbar v_F}$ and
$q_0(\epsilon) \approx k_F + \tfrac{\epsilon}{\hbar v_F}$, respectively.  To the
leading order in the semiclassical approximation, the off-diagonal elements of
the T-matrix in Eq~\eqref{eq:T-matrix-elements} are negligible and the diagonal
terms are given by $T_{++}^\sigma \approx e^{\sigma i\Phi}$, where
$\Phi \equiv \tfrac{hL}{\hbar v_F}$ is referred to as the magnetic phase.  This
expression for the T-matrix has a simple physical interpretation: within the
semiclassical approach the incoming electrons have an energy of the order of
$\mu$, much larger than the scattering potential height, $h$.  Hence, incoming
particles have a unit probability to be transmitted through the F region.
Propagation through the F region results only in the additional phase $\Phi$.
Clearly, the spectrum obtained from Eq~\eqref{eq:quant-condition} in this limit
coincides with the result of the Bohr-Sommerfeld quantization condition for the
spectrum:
\begin{equation}
  \label{eq:andreev-bound-states}
  \frac{\epsilon L}{\hbar v_F} + \sigma\Phi -
  \arccos \frac{\epsilon}{\Delta} \pm \frac{\varphi}{2} = \pi n,
\end{equation}
where $n$ is an integer. Eq. \ref{eq:andreev-bound-states}
determines the spectrum of Andreev bound states (ABS)
\cite{konschelle-2016-ballistic,konschelle-2016-semiclassical}. In a short
junction, $L \ll \xi_0$, where $\xi_0 \equiv \hbar v_F/\Delta$ is the
superconducting coherence length, one obtains $\epsilon_\sigma=\pm\Delta\cos\big(\sigma\Phi +
\tfrac{\varphi}{2}\big)$. It follows that  by changing the magnetic phase, the energy of
the ABS can be tuned between $\pm \Delta$. In particular, zero energy single
states can be created by proper choice of $\Phi$ and $\varphi$.
 
The other widely studied limiting case is the YSR limit in which the F region is
described by a $\delta$-like potential, i.e. its length tends to zero,
$L \rightarrow 0$, while $\Phi$ is kept finite. One can read directly from
Eqs~\eqref{eq:T-matrix-elements} that, within this limit,
$T_{++}^\sigma \approx 1 + \sigma i \Phi$ and the off-diagonal elements are
non-zero, $T_{+-}^\sigma \approx \sigma i \Phi$. This means that, in the
presence of a $\delta$-like potential, the backscattering probability is finite.
The latter leads to a coupling between the $\pm k_F$ valleys (see sketch in
Fig~1b). Such coupling lifts the degeneracy at $\varphi=0$ and "pushes" one of
the states to energies closer to the continuum.  By solving
Eq~\eqref{eq:quant-condition} in this limit for a general value of $\varphi$,
one obtains four bound states \cite{costa-2018-connection}:
\begin{align}
  \label{eq:YSR-bound-states}
  \epsilon = \pm \frac{\Delta}{\Phi^2+1} \bigg[&\Phi^4 + \frac{1-\cos\varphi}{2}\Phi^2 + \frac{1+\cos\varphi}{2} \nonumber \\
  & \pm \Phi \sqrt{2\Phi^2 (1+\cos\varphi) + \sin^2\varphi }\ \bigg]^{1/2}\; .
\end{align}
Here the $\pm$ signs are mutually independent and the bound states have to appear 
inside the gap,   $|\epsilon| \leq \Delta$. For a zero phase-difference,  $\varphi = 0$, 
there are only two states inside the gap, which correspond to the well-known YSR solution:
\begin{equation}
  \epsilon = \pm \Delta \dfrac{1 - \Phi^2}{1 + \Phi^2} \; .
\end{equation}
The other  two states  remain at the gap edges, $\epsilon = \pm \Delta$,
independently of the value of $\Phi$. Whereas the YSR are nondegenerate, ABS
states, Eq~\eqref{eq:andreev-bound-states}, are double degenerate.  Moreover,
with increasing $\Phi$ the ABS cross zero energy every time
$\Phi=(2n+1)\pi/2$. In contrast, YSR states cross the zero only once at
$\Phi=1$, where, as explained below, a quantum phase transition takes place
\cite{sakurai-1970-comments,balatsky-2006-impurity,franke-2011-competition}.

\subsection{Spectrum in an intermediate range of parameters }
\label{sec:spectral-properties}

\begin{figure}[t!]
  \centering
  \includegraphics[width=\linewidth]{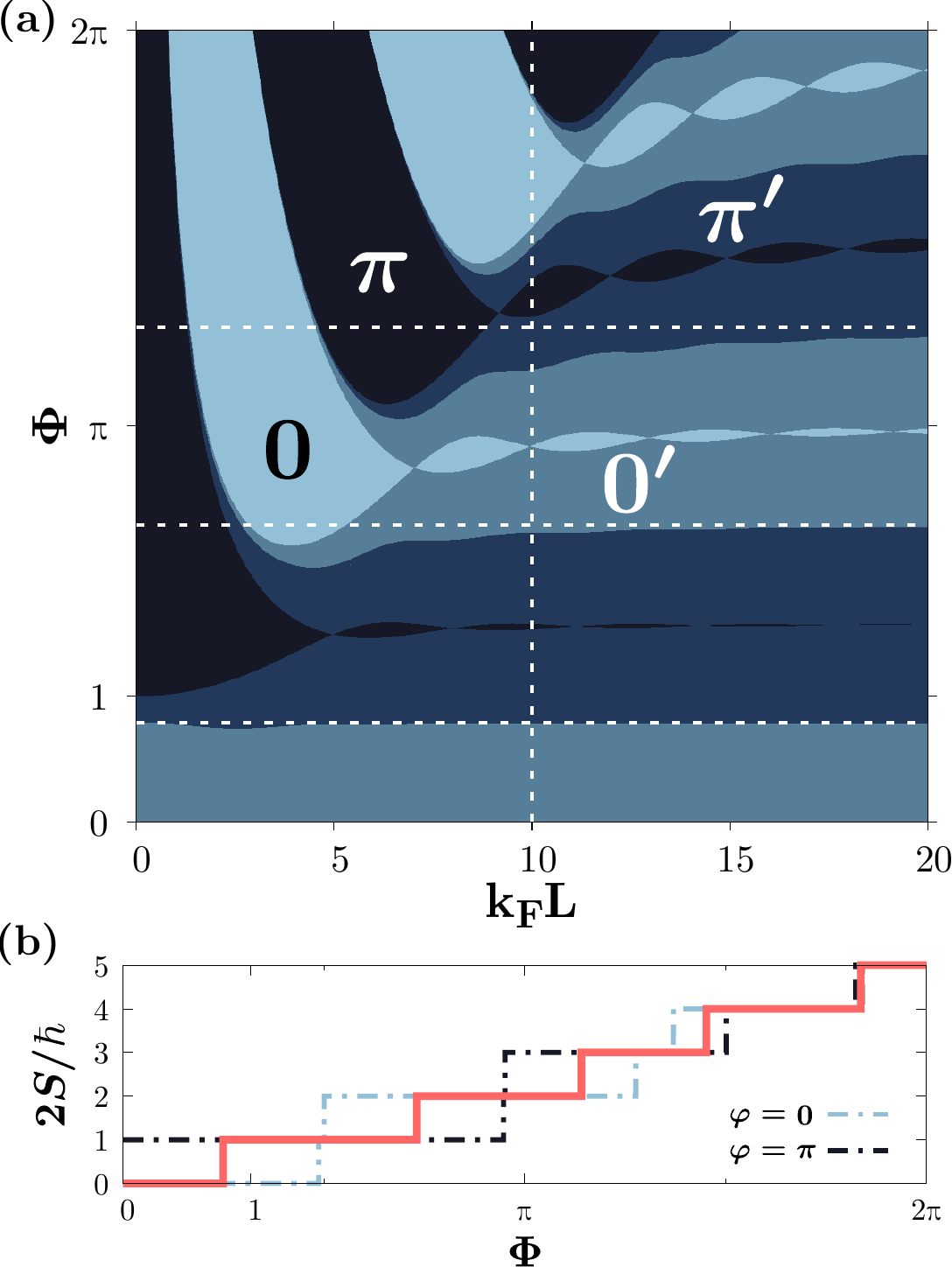}
  \caption{(a) Phase diagram of the SFS Josephson junction in terms of the
    length of the junction $L$ and the magnetic phase $\Phi$. The horizontal
    white dashed lines indicate the values of $\Phi$ chosen in
    Fig~\ref{fig:bound-states}. (b) Total electronic spin of a SFS junction of
    length $k_FL=10$ (white dashed line in panel a) when one imposes $0$-phase
    (dashed light) or $\pi$-phase (dashed dark).  The red solid line shows the
    total spin when the junction stays in its ground state. Calculations have
    been done for $\mu/\Delta = 100$.}
  \label{fig:phase-diagrams}
\end{figure}

We address now the question about the spectrum in an intermediate case, between the semiclassical and the YSR limits. 
 This may correspond to  a cluster of magnetic atoms or a small ferromagnetic island
with a large but finite exchange field. The expression determining the bound states 
 can be obtained from Eqs~\eqref{eq:quant-condition} and
\eqref{eq:T-matrix-elements} and it is explicitly shown in the appendix, 
Eq~\eqref{eq:app-full-expression-quantization}. In Fig~\ref{fig:bound-states}
we show with solid lines the subgap spectrum of the SFS structure as a function
of the normalized length of the magnetic region, $k_F L$, for $\varphi=0$
. Different panels correspond to different values of the magnetic phase
$\Phi$. For small $k_F L\lesssim1$ there are only two nondegenerate states
within the gap. These are the YSR states. Figures~\ref{fig:bound-states}a and
\ref{fig:bound-states}b correspond, respectively, to the situations before and
after the YSR states cross at zero energy.  Further increase of $\Phi$ pushes
the states towards the gap edges. In contrast, for longer junctions,
$k_F L\gg 1$, two pairs of bound states can be found within the gap. These pairs
of states are non-degenerate (except at certain values of $k_F L$) and their
energy oscillates with a period $2\pi/k_F$ around the semiclassical value
determined by Eq~\eqref{eq:andreev-bound-states}. The oscillations stem from
interference effects that are ignored in the semiclassical limit. Further
increase of the junction length  towards $L\sim \xi_0$ will bring additional bound
states into the gap, which are not considered here.

It is worth noticing that Figs~\ref{fig:bound-states}b-c show zero-energy
crossings for finite length junctions at $\varphi=0$.  At each crossing the
total spin of the system change by one, as calculated from
Eq~\eqref{eq:magnetization} and shown by dashed black lines in
Fig~\ref{fig:bound-states}. In other words, Fig~\ref{fig:bound-states}
demonstrates that a QPT also takes place beyond the YSR limit. Moreover, a
sequence of QPTs with a stepwise change of the total spin may exist in a finite
length junction.

The number of zero-energy crossings as a function of $L$ grows with increasing
$\Phi$ \footnote{See multimedia material in the supplementary information}. As
it follows from Eq~\eqref{eq:andreev-bound-states}, in a short junction within
the semiclassical limit, $k_F^{-1} \ll L \ll \xi_0$, the ABS cross zero
periodically at values of the magnetic phase $\Phi=(2n+1)\pi/2$.  Each of these
``asymptotic'' crossings should be accompanied with, at least, two additional
zero-energy crossings at intermediate values of $k_F L$
(Fig~\ref{fig:bound-states}b-c). Fast oscillations of the bound state energies
as a function of $L$ may increase further the number of zero-energy crossings by
an even number (Fig~\ref{fig:bound-states}c).

\subsection{Josephson current}
\label{sec:josephson-current}

The subgap spectrum can be measured by means of  tunneling spectroscopy
\cite{yazdani-1997-probing,ji-2008-high,pillet-2010-andreev,ruby-2015-tunneling,
  heinrich-2018-single}. In addition, measurements of the Josephson current in SFS junctions
can also shed light on the spectral properties
\cite{goffman-2000-supercurrent,agrait-2003-quantum}, in particular on the
ground state of the junction. In conventional SNS junctions the Josephson energy
is minimized when the phase difference vanishes, $\varphi=0$. However, it is
known that, in SFS junctions, this minimum can also be found at $\varphi=\pi$ by
tuning the exchange field or the length of the F region
\cite{bulaevskii-1977-superconducting,buzdin-1982-critical,lazar-2000-superconductor,ryazanov-2001-coupling,chtchelkatchev-2001-pi0}. In
the context of a delta-like magnetic impurity the connection between the zero
energy YSR state and the $0$-$\pi$ transition has been recently discussed in
Ref~\cite{costa-2018-connection}.  As we discuss next, the transition between
the $0$- and $\pi$-junction behaviour is closely related to the QPTs described
above for arbitrary junctions.
  
For this sake, we compute the ground state energy of the junction as a function
of the phase difference $\varphi$.  If the energy has a unique minimum at
$\varphi=0$ or $\varphi=\pi$, one says that the junction is in the 0- or $\pi$-
phase respectively.  If the Josephson energy has minima both at $\varphi=0$ and
at $\varphi=\pi$, then the ground state is denoted as $0'$ or $\pi'$ depending
on the location of the global minimum \cite{rozhkov-1999-josephson,
  vecino-2003-josephson,bergeret2006interplay}.

In Fig~\ref{fig:phase-diagrams}a we show the phase diagram in the $L$-$\Phi$
plane. This diagram provides an interesting connection: the QPTs associated with
the zero-energy crossings shown in Figs~\ref{fig:bound-states}(b-c) (horizontal
dashed lines in Fig~\ref{fig:phase-diagrams}a), correspond to transitions
between the ($0$, $0'$, $\pi'$) states and the $\pi$ state.

Finally, in Fig~\ref{fig:phase-diagrams}b we show the dependence of the total spin of the system
on $\Phi$ for a junction with $k_FL=10$.  The dashed light line (dashed dark line)
shows the spin if the junction is forced to stay in the $0$($\pi$)-state. The
solid red line shows the spin of the system if the junctions always stays in the
true ground state, i.e., if it follows the global  energy minimum when the parameters are
changed. Notice that, whenever the ground state corresponds to $\varphi = 0$
($\varphi = \pi$), the total electronic spin of the system is even (odd).

\section{Conclusions}
\label{sec:summary}

In conclusion, we present a complete study of equilibrium properties of a
superconducting wire with a magnetic defect.  We derive a general expression, Eq
\eqref{eq:quant-condition}, that determines the full subgap spectrum provided
that the T-matrix of the F region in the normal state is known.  We also
demonstrate in Eqs~\eqref{eq:magnetization} and
\eqref{eq:magnetization-zero-temp} that the total spin of a SFS junction
undergoes integer jumps in units of $\hbar/2$ associated with zero-energy
crossings of the bound states. Specifically, we analyze the spectrum of a
one-dimensional ballistic SFS Josephson with a F region smaller than the
superconducting coherence length but arbitrary strength of the exchange
field. Our theoretical analysis bridges nicely two previously disconnected
limiting cases: the YSR and the semiclassical ones.  We demonstrate that the QPT
predicted by the YSR model can be also found for SFS junctions of finite length
$L$. Such phase transitions are associated not only to the integer jumps of the
total spin described by our generalized Friedel sum rule, but also to a change
of the sign of the supercurrent.

\section*{Acknowledgments}

We acknowledge funding by the Spanish Ministerio de
Econom\'ia y Competitividad (MINECO) (Projects No. FIS2014-55987-P,
FIS2016-79464-P and FIS2017-82804-P).  I.V.T. acknowledges support by Grupos
Consolidados UPV/EHU del Gobierno Vasco (Grant No. IT578-13).  M.R and
F.S.B. acknowledge funding from the EU's Horizon 2020 research and innovation
programme under grant agreement No. 800923 (SUPERTED).

\appendix{}

\section{T-matrix of the F region}
\label{sec:app-T-matrix}

Here we derive the normal state T-matrix of a ferromagnetic region of length $L$ and Zeeman
splitting $h$ centered at the origin between two metallic electrodes. This matrix enters 
 Eq \eqref{eq:quant-condition} and hence it is pivotal to obtain the
 the bound states. In the normal state electrons and holes are
decoupled, so we will only focus on the electrons. The wavefunction reads
\begin{equation}
  \label{eq:NFN-wavefunction}
  \psi (x) = \left\{
    \begin{array}{lcl}
      A_L^\sigma e^{iq_0x} + B_L^\sigma e^{-iq_0x} & \text{if} & x < -L/2 \\
      C^\sigma e^{iq_\sigma x} + D^\sigma e^{-iq_\sigma x} & \text{if} & -L/2 < x < L/2 \\
      A_R^\sigma e^{iq_0x} + B_R^\sigma e^{-iq_0x} & \text{if} & x > L/2
    \end{array}
    \right. \ ,
\end{equation}
where $q_\sigma = k_F\sqrt{1 + \frac{\epsilon+\sigma h}{\mu}}$ and
$q_0 = k_F\sqrt{1 + \frac{\epsilon}{\mu}}$ are the wavenumbers at the
ferromagnet and the normal metal, $k_F$ is the Fermi wavenumber and $\sigma=\pm$
stands for the spin orientation. From the continuity of the wavefunction in Eq
\eqref{eq:NFN-wavefunction} and its first derivative, we obtain a set of four
equations that we have to solve. First writing $C^\sigma$ and $D^\sigma$ in
terms of $A_L^\sigma$ and $B_L^\sigma$, and substituting them into the
expressions for write $A_R$ and $B_R$, we finally get a connection between the
wavefunction at the left and right superconductor,
\begin{equation}
  \label{eq:NFN-matricial}
  \left(
    \begin{array}{c}
      A_R^\sigma \\ B_R^\sigma
    \end{array} \right)
  =
  \left(
    \begin{array}{cc}
      T_{11}^\sigma & T_{12}^\sigma \\
      T_{21}^\sigma & T_{22}^\sigma
    \end{array}\right)
  \left(
    \begin{array}{c}
      A_L^\sigma \\ B_L^\sigma
    \end{array}\right),
\end{equation}
where
\begin{align}
  \label{eq:NFN-T11}
  &T_{11}^\sigma = \bigg[
    \cos\Big( q_{\sigma} L \Big) +
    \frac{i}{2}\frac{{q_{\sigma}}^2 + {q_0}^2}{q_{\sigma} q_0}
    \sin \Big(q_{\sigma} L\Big)
    \bigg] e^{-iq_0 L}, \\
  \label{eq:NFN-T12}
  &T_{12}^{\sigma} = \frac{i}{2}\frac{{q_{\sigma}}^2 - {q_0}^2}{q_{\sigma} q_0}
    \sin \Big(q_{\sigma} L\Big)\; ,
\end{align}
and the remaining two components are related to these ones by complex
conjugation, $T_{22}^\sigma = (T_{11}^\sigma)^*$ and
$T_{21}^\sigma = (T_{12}^\sigma)^*$. As it is defined in Eq
\eqref{eq:L-to-R-connection-electrons}, the matrix in Eq
\eqref{eq:NFN-matricial} is the normal state transfer matrix of the
ferromagnetic region.

\section{The S-F-S subgap spectra}
\label{sec:app-subgap-spectra}

Here we obtain the spectrum of a homogeneous S-F-S junction, whatever the values
of the width and the exchange field strength of the magnetic region are. We
start from the secular equation \eqref{eq:quant-condition} and assume that
$\hat T^\sigma$ is a generic $2 \times 2$ matrix, like the one in Eq
\eqref{eq:NFN-matricial}. After some algebra and exploiting the relations
between the elements of the transfer matrix, we get a rather simple equation
\begin{equation}
  \label{eq:quant-condition-developed}
  \cos \varphi - \text{Re} \bigg[
  T_{12}^\sigma \overline T_{12}^\sigma + e^{2i\alpha} T_{11}^{\sigma^*} \;  \overline T_{11}^\sigma
  \bigg] = 0,
\end{equation}
from which, substituting the expressions for the elements of the T-matrix in Eqs
\eqref{eq:NFN-T11} and \eqref{eq:NFN-T12}, we obtain
\begin{widetext}
\begin{align}
  2\cos\varphi & - 2\cos(2\alpha) \cos(q_{\sigma} L) \cos(\overline{q_{\sigma}} L)
  - \frac{k_F^2}{q_{\sigma} \overline{q_{\sigma}}} \bigg[
  2cos(2\alpha) + \Big(\tfrac{\epsilon+\sigma h}{\mu}\Big)^2 \sin^2\alpha
                 \bigg] \sin(q_{\sigma} L) \sin(\overline{q_{\sigma}} L) \nonumber\\
  \label{eq:app-full-expression-quantization}
               & - \frac{{q_{\sigma}}^2 + k_F^2}{q_{\sigma} k_F} \sin(2\alpha) \sin(q_{\sigma} L) \cos(\overline{q_{\sigma}} L)
                 + \frac{\overline{q_{\sigma}}^2 + k_F^2}{\overline{q_{\sigma}} k_F} \sin(2\alpha) \cos(q_{\sigma} L) \sin(\overline{q_{\sigma}} L)= 0.
\end{align}  
\end{widetext}

In Eq \eqref{eq:app-full-expression-quantization},
$\overline{q_\sigma} = k_F\sqrt{1-\frac{\epsilon + \sigma h}{\mu}}$ is the time
conjugate of the electron wavenumber in F and we have approximated
$q_0 \approx k_F$, which is totally justified by the fact that $\Delta \ll \mu$
is fulfilled in any superconductor and that $q_0$ did not appear in any
trigonometric function (where the accumulated phases along long distances would
eventually be non-negligible, $\epsilon/\mu \cdot k_FL \sim 2\pi$).

\bibliographystyle{apsrev4-1}
\bibliography{list.bib}

\end{document}